\documentclass[letterpaper,12pt]{article}

\usepackage{cite}
\usepackage{epsfig}
\usepackage[english]{babel}
\usepackage{amsmath}
\usepackage{amssymb}
\usepackage{amsbsy}
\usepackage{amstext}
\usepackage{amscd}
\usepackage{amsxtra}
\usepackage{amsopn}
\usepackage[onehalfspacing]{setspace}
\usepackage{mathrsfs}
\usepackage{relsize}
\usepackage{amsfonts}
\usepackage[makeroom]{cancel}
\usepackage{selinput}

\usepackage{graphics}
\usepackage[export]{adjustbox}
\usepackage{subfig}

\newcommand{\dd}{\mbox{d}}

\newcommand{\ts}[1]{{\boldsymbol{#1}}}

\usepackage{geometry}
\usepackage{parskip}

\usepackage[colorlinks=true,citecolor=green,linkcolor=red,filecolor=cyan,urlcolor=magenta,backref=page]{hyperref}
\renewcommand*{\backref}[1]{}
\renewcommand*{\backrefalt}[4]{%
    \ifcase #1 (Not cited.)%
    \or        (Cited on page~#2.)%
    \else      (Cited on pages~#2.)%
    \fi}

\usepackage{color}
\definecolor{mygray}{rgb}{0.5,0.5,0.5}

\begin{document}

\title{Non-singular ``Gauss'' black hole from non-locality}

\author{Jens Boos\,\footnote{~E-mail: \href{mailto:jens.boos@kit.edu}{jens.boos@kit.edu}} \\
    {\small High Energy Theory Group, Department of Physics, William \& Mary}\\[-8pt]
    {\small Williamsburg, VA 23187-8795, United States} \\
	{\small Institute for Theoretical Physics, Karlsruhe Institute of Technology}\\[-8pt]
    {\small D-76128 Karlsruhe, Germany} }

\date{Mar 10, 2025}

\maketitle

\begin{abstract}

Cutting out an infinite tube around $r=0$ formally removes the Schwarzschild singularity, but without a physical mechanism this procedure seems ad hoc and artificial. In this paper we provide justification for such a mechanism by means of non-locality. Motivated by the Gauss law we define a suitable radius variable as the inverse of a regular non-local potential, and use this variable to model a non-singular black hole. The resulting geometry has a de\,Sitter core, and for generic values of the regulator there is \emph{no inner horizon}, saving this model from potential issues via mass inflation. An \emph{outer} horizon only exists for masses above a critical threshold, thereby reproducing the conjectured ``mass gap'' for black holes in non-local theories. The geometry's density and pressure terms decrease exponentially, thereby rendering it an almost-exact vacuum solution of the Einstein equations outside of astrophysical black holes. Its thermodynamic properties resemble that of the Hayward black hole, with the notable exception that for critical mass the horizon radius is zero.

\end{abstract}

\vfill

\pagebreak

\section{Introduction}

The presence of singularities inside black holes is a robust prediction of General Relativity. However, it is commonly believed that a suitable UV completion of gravity ameliorates this behavior and renders all physical quantities finite in proximity to the center of the black hole. While there are indications that putative theories of quantum gravity feature regular black holes in their semiclassical limits, an explicit derivation of such objects proves cumbersome.

For this reason, Bardeen \cite{Bardeen:1968} considered a simple modification of the Schwarzschild metric that is manifestly finite at $r=0$ but reproduces the large-distance behavior known from General Relativity. Others have followed similar approaches and have developed a rich family of non-singular black hole geometries \cite{Dymnikova:1992ux,Bonanno:2000ep,Hayward:2005gi,Frolov:2014jva,Frolov:2016pav,Frolov:2017rjz,Cano:2018aod,Simpson:2018tsi,Nicolini:2019irw,Simpson:2019mud,Bonanno:2020fgp} (and references therein). In this paper we focus on static regular black holes and postpone a discussion of time-dependent formation (and evaporation) to later studies. Static non-singular black hole geometries typically have several properties:
\begin{enumerate}
\item They do not solve the vacuum Einstein equations exactly, but their Einstein tensor decreases polynomially with distance away from the center at $r=0$. Alternatively, this can be viewed as the presence of an effective energy-momentum tensor, and the properties of this matter source can be analyzed with respect to energy conditions. In accordance with Penrose's singularity theorem, an energy condition is violated if the inner black hole singularity is avoided.
\item In addition to the outer event horizon at $r \approx 2GM$ there exists an inner horizon at $r \sim \ell$ as well, where $\ell$ is the regularization scale.
\item Close to $r=0$ the geometry approaches a de\,Sitter form.
\item The curvature upper bound is given by $1/\ell^2$ and is independent of the black hole mass, which is also called the ``limiting curvature condition'' \cite{Markov:1982,Markov:1984,Polchinski:1989,Frolov:2016pav}.
\item At large distances $r \ll \ell$, the regulator terms decrease rapidly and the metric increasingly approximates the Schwarzschild metric of general relativity.
\end{enumerate}
Moreover, in the spherically symmetric and static case, the regularity is achieved by replacing the mass parameter $M$ by a mass function $M(r)$ that scales in a suitable fashion to remove the singularity at $r=0$. A well-known model is that of Hayward \cite{Hayward:2005gi},
\begin{align}
\dd s^2 = -f_H(r)\dd t^2 + \frac{\dd r^2}{f_H(r)} + r^2 \dd\Omega^2 \, , \quad f_H(r) = 1 - \frac{2Mr^2}{r^3+2M\ell^2} \, ,
\end{align}
where $\ell > 0$ is the regularization length scale and we employ units wherein $G=1$. The complicated appearance of the black hole mass parameter $M$ in the denominator of the function $f_H(r)$ guarantees the limiting curvature condition. Typically, the function $f_H(r)$ has two zeroes, corresponding to the inner horizon and the outer horizon, respectively.

Of course, in the absence of a fundamental theory predicting the precise form of a non-singular metric, many different parametrizations can be explored. From a fundamental physics perspective, however, this is somewhat dissatisfying, since there is no physical argument that favors one type of non-singular metric over another, equally non-singular one. In this paper we propose an avenue to approach this problem by connecting the regularity properties of static black hole spacetime metrics with Gauss' law. Starting from a modified radius variable we construct a non-singular metric that turns out to not have an inner horizon but still features a de\,Sitter core. The form of the modified radius variable is motivated by recent results in non-local gravity, thereby removing a layer of ambiguities.

\section{Modified radius variable}
\begin{figure}[!htb]
\centering
\subfloat{ \includegraphics[width=0.5\textwidth]{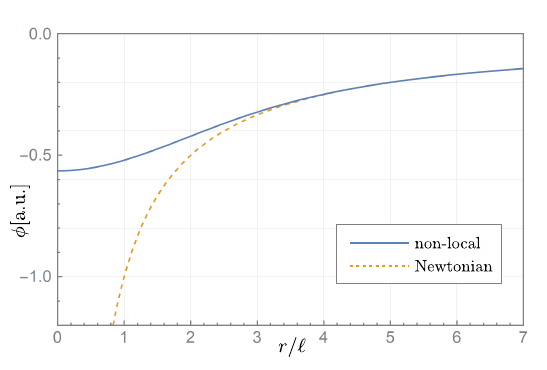} }
\subfloat{ \includegraphics[width=0.5\textwidth]{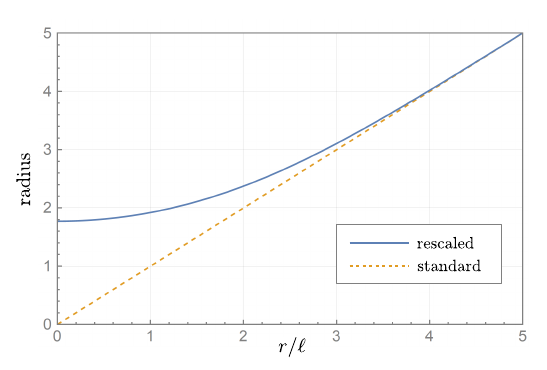} }
\caption{Newtonian and regularized potential (left), and corresponding radius functions (right).}
\label{fig:1}
\end{figure}

In a local field theory in four spacetime dimensions, the potential of a point particle decreases monotonically with the inverse spatial distance (in suitable units),
\begin{align}
\phi_\text{loc} = -\frac{1}{r} \, .
\end{align}
Similarly, the field strength decreases with the inverse area, due to Gauss' law. Simply speaking, this is a consequence of the Poisson equation,
\begin{align}
\nabla^2 \phi_\text{loc}(r) = -4\pi \, \delta{}^{(r)}(\ts{r}) \, .
\end{align}
Now, reversing this logic, one could measure the field strength and thereby deduce the radial distance away from the source. As the field strength diverges, one reaches $r=0$. For the sake of simplicity, but without loss of generality, in what follows we shall consider the potential as the fundamental variable, for which similar mathematical properties hold true. Hence, one may be inclined to define a radius to be the inverse of the potential,
\begin{align}
r \equiv -\frac{1}{\phi_\text{loc}} \, .
\end{align}
However, the singularity of the local potential is deemed unphysical since it gives rise to infinite forces and accelerations. It is possible to modify the equations of motion for scalar potentials, and at the linear level a class of non-local theories has proven particularly successful in removing the divergence at $r=0$ \cite{Tomboulis:1997gg,Biswas:2005qr,Modesto:2011kw,Biswas:2011ar,Edholm:2016hbt,Giacchini:2018wlf,Boos:2018bxf,Boos:2020qgg}; for earlier work in non-commutative geometry and regular black holes see Refs.~\cite{Nicolini:2005zi,Nicolini:2005vd,Spallucci:2006zj,Modesto:2010uh}. Within a quantum-mechanical approach to the singularity problem one also encounters non-local terms \cite{Greenwood:2008ht,Saini:2014qpa}. For these reasons we consider the non-local equation
\begin{align}
\label{eq:non-local}
F(\nabla^2) \nabla^2 \phi_\text{nl} = -4\pi \, \delta{}^{(3)}(\ts{x}) \, .
\end{align}
Therein, $F(\nabla^2)$ is a so-called \emph{form factor} that depends on a regularization parameter $\ell > 0$ and that satisfies $F(0) = 1$. A popular choice motivated from string field theory is \cite{Biswas:2005qr}
\begin{align}
\label{eq:gf-form-factor}
F(\nabla^2) = e^{-\ell^2\nabla^2} \, .
\end{align}
This equation can be used with the method of non-local Green functions; for a comprehensive review we refer to Ch.~2.8 in \cite{Boos:2020qgg} as well as references therein. For a general form factor, the spherically symmetric point particle solution takes the form
\begin{align}
\phi_\text{nl}(r) = - \frac{1}{r} \sqrt{\frac{2}{\pi}} \int\limits_0^\infty \frac{\dd z}{\sqrt{z}} \, \frac{1}{F\left(-\frac{z^2}{r^2}\right)} J_{1/2}(z) \, ,
\end{align}
where $J_{1/2}(z) = \sqrt{2/(\pi x)}\sin x$ is the Bessel function of the first kind, and the Coulomb potential is recovered by setting $F \equiv 1$ and using $\int_0^\infty \dd x \, x^{-1}\,\sin x = \pi/2$. For the above choice of form factor one readily obtains
\begin{align}
\phi_\text{nl} = -\frac{\text{erf}\left(\frac{r}{2\ell}\right)}{r} \, ,
\end{align}
where $\text{erf}(x)$ denotes the error function which asymptotes exponentially fast to unity \cite{Olver:2010},
\begin{align}
\text{erf}(x\rightarrow\infty) \approx 1 - \frac{e^{-x^2}}{\sqrt{\pi} x} \, .
\end{align}
This relation guarantees that in the limit $r/(2\ell) \rightarrow \infty$ (that is, at large distances $r$ compared to the regulator $\ell$, or at vanishing regulator scale $\ell \rightarrow 0$ at fixed $r$) we recover the Coulomb potential. At small distances, however, this potential differs appreciably from the singular Coulomb potential: it is finite and regular at $r=0$. Using this non-locally regularized potential we may now define a modified radial distance
\begin{align}
\tilde{r} \equiv -\frac{1}{\phi_\text{nl}} = \frac{r}{\text{erf}\left(\frac{r}{2\ell}\right)} \, .
\end{align}
In Fig.~\ref{fig:1} we plot the local potential with its regularized, non-local counterpart, as well as the two corresponding radius variables. As becomes apparent, the modified radius variable $\tilde{r}$ has a minimal value proportional to the regulator scale $\ell$:
\begin{align}
\tilde{r}(r \rightarrow 0) = \sqrt{\pi}\ell + \mathcal{O}(r^2) \, .
\end{align}
At large distances, however, the two radial coordinates approach each other exponentially fast \cite{Olver:2010},
\begin{align}
\tilde{r}(r \rightarrow \infty) = r + \frac{2\ell}{\sqrt{\pi}} e^{-r^2/(4\ell^2)} \, .
\end{align}
Hence, taking this non-locally modified radius variable $\tilde{r}$ as the physical radius variable effectively \emph{cuts out} the region $r\in[0,\sqrt{\pi}\ell]$ from the manifold, while rapidly approaching the standard radius definition for distances larger than $\ell$.

\section{Non-singular ``Gauss'' black hole model}
Let us now explore the ramifications for a static and spherically symmetric black hole spacetime subjected to the formal substitution $r \mapsto \tilde{r}(r)$,
\begin{align}
\dd s^2 = -f_\text{nl}(r)\dd t^2 + \frac{\dd r^2}{f_\text{nl}(r)} + \tilde{r}^2 \dd\Omega^2 \, , \quad f_\text{nl}(r) = 1 - \frac{2M}{\tilde{r}(r)} \, , \quad \tilde{r} = \frac{r}{\text{erf}\left(\frac{r}{2\ell}\right)} \, .
\end{align}
Due to its motivation via the non-local Gauss law (and the appearance of the error function $\text{erf}(x)$ as well as Gaussian factors $e^{-r^2/(4\ell^2)}$ in the radius and curvature) we shall refer to it as the ``Gauss'' model. Note that this is not a coordinate transformation since we explicitly keep $r$ as the coordinate radius variable. However, it is clear that circles of $r=\text{const}$ now have the proper circumference $2\pi\tilde{r}(r)$. Unlike usually assumed in non-singular black hole models, we here explicitly rescale the spherical part of the geometry as well, which is a necessary step to render this black hole model finite at $r=0$. This is similar to the model proposed by Simpson and Visser \cite{Simpson:2018tsi}. In what follows, we will discuss this metric in more detail. In particular,we will discuss (i) the horizons, and, in particular, the absence of an inner horizon; (ii) the absence of an outer horizon for large regulators (``mass gap''); (iii) a thorough study of curvature invariants including the Kretschmann scalar as well as squared of the Weyl tensor, the tracefree Ricci tensor, and the Ricci scalar, demonstrating manifest regularity of this metric; (iv) the question of universal boundedness in curvature (``limiting curvature condition''); (v) the properties of the effective energy-momentum tensor and violation of energy conditions; (vi) the Hawking temperature and entropy of this metric as compared to the Schwarzschild case, and, finally, (vii) the interpretation of the hypersurface $r=0$ in relation to wormholes and geodesic completeness.

\subsection{Horizons}

\begin{figure}[!htb]
\centering
\includegraphics[width=0.7\textwidth]{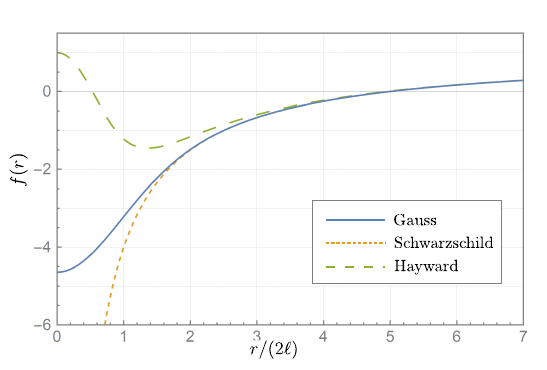}
\caption{Metric functions for the Schwarzschild, Hayward, and Gauss black hole.}
\label{fig:2}
\end{figure}

Let us briefly compare the metric function $f_\text{nl}(r)$ to that of General Relativity and Hayward, see Fig.~\ref{fig:2}. For generic values of $\ell$ and $M$, where we assume that $M/\ell > 1$, it is clear that the behavior at $r=0$ is rather different. In the General Relativity case one has the standard spacelike singularity, whereas the Hayward model is de\,Sitter-like. At $r=0$ the Gauss model behaves as
\begin{align}
f(r\ll \ell) \approx 1 - \frac{2M}{\sqrt{\pi}\ell} + \frac{M r^2}{6\sqrt{\pi}\ell^3} \, ,
\end{align}
which shows that for large masses $2M > \sqrt{\pi}\ell$ the geometry is indeed de\,Sitter-like at the origin.

\begin{figure}[!htb]
\centering
\includegraphics[width=0.7\textwidth]{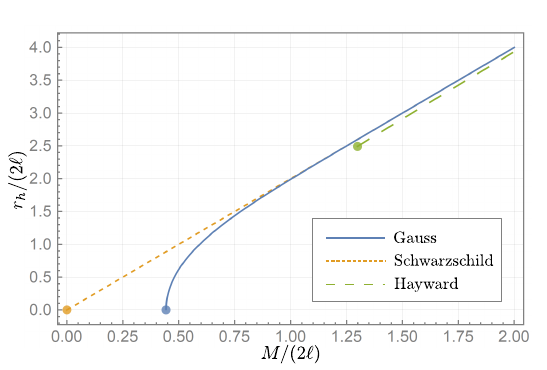}
\caption{Horizon radii for the Gauss black hole compared to the Schwarzschild metric and Hayward metric. Note that the horizon radius for the Gauss black hole at critical mass $M_0$ is zero.}
\label{fig:3c}
\end{figure}

The striking difference between the Gauss and Hayward model lies in the \emph{absence of an inner horizon} for the latter. An apparent horizon is located wherever the following condition is satisfied:
\begin{align}
\label{eq:app-horizon}
(\nabla r)^2 = g^{rr} = 0 \, ,
\end{align}
such that the locations of apparent horizons correspond to the zeros of the metric function $f(r)$, or, equivalently, wherever the vector field $\partial^\mu r = \delta^\mu_r$ becomes null. While the outer horizons are roughly located around $r \sim 2M$, modulo small corrections due to $\ell$, there is an inner horizon for the Hayward model, but none for the Gauss and Schwarzschild black hole; see Fig.~\ref{fig:3c} for a visualization of horizon radii.

Since inner horizons make black holes susceptible to mass inflation \cite{Poisson:1989zz,Poisson:1990eh}, the generic absence of such a structure in this model is an interesting feature of the non-local regulator. While more work is needed to understand the precise origins, it is likely due to the fact that our model is intrinsically non-polynomial. In this way, the absence of the inner horizon would be directly inherited from the ``ghost-free property'' of non-local gravity which in turn heavily relies upon entire non-polynomial functions for the gravitational propagator \cite{Biswas:2013kla,Buoninfante:2018mre,Boos:2020qgg}, such as $e^{-\ell^2\nabla^2}$ as employed in Eq.~\eqref{eq:non-local}.

\begin{figure}[!htb]
\centering
\subfloat{ \includegraphics[width=0.5\textwidth]{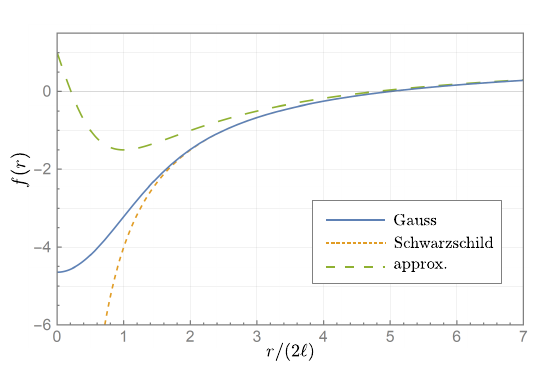} }
\subfloat{ \includegraphics[width=0.5\textwidth]{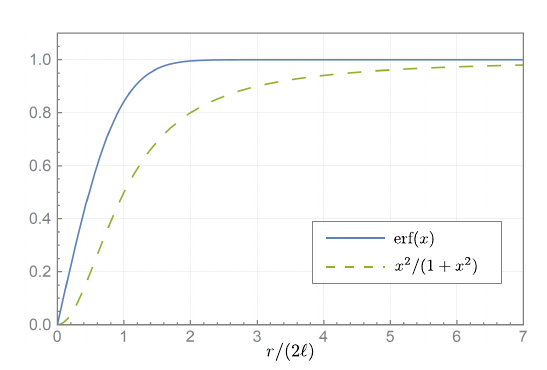} }
\caption{Left: The metric function $f(r)$ for the Gauss regular black hole (solid line), the Schwarzschild metric (dotted line), and the Gauss regular black hole subjected to the approximation $\text{erf}(x) \rightarrow x^2/(1+x^2)$ (dashed line). Clearly, this approximation induces an inner horizon. Right: The error function $\text{erf}(x)$ and its approximation $x^2/(1+x^2)$ in direct comparison.}
\label{fig:3}
\end{figure}

In fact, one may check that substituting the complicated function $\text{erf}(x)$ by a rational approximation $x^2/(1+x^2)$ gives rise to an inner horizon; see Fig.~\ref{fig:3} for a plot of the metric functions as well as the error function and its approximation. (The substitution $x/(1+x)$ is not allowed since it induces a conical singularity around $r=0$.) For this reason we believe that the absence of the inner horizon is indeed due to the non-rational form of our modification.

Let us understand the consequences of the error function approximation at a deeper level. To that end, in non-local field theory we may write the inverse of the form factor as the following regularized integral over the potential of a point particle \cite{Boos:2020qgg},
\begin{align}
\frac{1}{F(-k^2)} = -k \lim_{\epsilon\rightarrow 0} \int\limits_0^\infty \dd r \, e^{-\epsilon r^2} \, r \, \sin(kr) \phi(r) \, .
\end{align}
One may verify that setting $\phi(r) = -1/r$ yields $F(-k^2) = 1$, as expected. A non-trivial check for $\phi(r) = \phi_\text{nl}(r)$, however, gives instead
\begin{align}
\frac{1}{F(-k^2)} = +k \lim_{\epsilon\rightarrow 0} \int\limits_0^\infty \dd r \, e^{-\epsilon r^2} \, r \, \sin(kr) \frac{\text{erf}\left(\frac{r}{2\ell}\right)}{r} = e^{-k^2\ell^2} \, ,
\end{align}
meaning $F(-k^2) = e^{k^2\ell^2}$, in exact correspondence to Eq.~\eqref{eq:gf-form-factor} under the Fourier substitution $\nabla^2 \rightarrow -k^2$. The propagator $\mathcal{D}$ of this theory is schematically given by
\begin{align}
\mathcal{D} \sim \frac{1}{k^2} \frac{1}{F(-k^2)} \, ,
\end{align}
This implies that every pole of this function corresponds to a propagating degree of freedom \cite{Buoninfante:2018mre}. The function $F(-k^2) = e^{k^2\ell^2}$ is everywhere non-vanishing, which implies that for non-local theories there are no additional propagating degrees of freedom. Incidentally, this is one of the reasons that non-local theories are sometimes also referred to as ``ghost-free.''

If one instead computes the above integral for the approximated error function, one finds
\begin{align}
\frac{1}{F(-k^2)} = 1 + (k\ell)\left[ e^{2k \ell} \text{Ei}(-2k\ell) - e^{-2k \ell} \text{Ei}(+2k\ell) \right] \, .
\end{align}
One can easily verify that $1/F(-k^2)$ assumes negative values for $k$ exceeding a critical value $k_0$ where $1/F(k_0^2) = 0$. This then makes the propagator of the theory change sign above a certain energy threshold, which has been shown to be related to instabilities \cite{Buoninfante:2018mre}, thereby demonstrating the pathological features of such a rational function approximation.

We note, finally, that an inner horizon also does not exist for Simpson--Visser choice \cite{Simpson:2018tsi}, which is given by $f(r) = 1 - 2M/\sqrt{r^2+\ell^2}$ and $\tilde{r} = \sqrt{r^2+\ell^2}$ and uses the square root function. Conversely, in Frolov's regular black hole models that utilize rational functions, an inner horizon persists \cite{Frolov:2016pav}. These considerations hence further underline the apparent necessity of non-rational regular black hole metrics if one wants to avoid an inner horizon.

\subsection{Mass gap}

It is well known that in higher-derivative as well as non-local infinite-derivative theories of gravity there exists a mass gap for the dynamical formation of black holes via a spherically symmetric collapse of null dust \cite{Frolov:2015bia,Frolov:2015bta}, and this mass gap is proportional to the regularization scale. In other words, small black holes do not form unless their mass parameter exceeds a critical value.

In the present context, note that the modified radius variable $\tilde{r}$ is always larger than the minimal distance $\sqrt{\pi}\ell$. For this reason the apparent horizon condition \eqref{eq:app-horizon} can only be satisfied if
\begin{align}
M > M_0 = \frac{\sqrt{\pi}\ell}{2} \, ,
\end{align}
that is, the mass parameter exceeds a critical value. As expected, in the limiting case of $\ell\rightarrow 0$ this mass gap vanishes as one recovers the Schwarzschild case. While the considerations presented in this paper are focused on the time-independent scenario, it is still interesting that they qualitatively reproduce the mass gap found in dynamical situations.

If the mass is less than the critical value, $M < M_0$, the resulting geometry is horizonless but regular at $r=0$. Specifically, $r=0$ then corresponds to a wormhole throat moving forward in time, just as in the Simpson--Visser case \cite{Simpson:2018tsi}; for comments on analytic continuation see below.

Let us conclude this section by addressing an interesting feature of the proposed ``Gauss'' regular black hole: at minimal mass, $M = M_0$, the horizon radius of the Gauss black hole is zero. Conversely, for the Hayward metric evaluated at its critical mass, the horizon radius is non-zero. This will become relevant in the thermodynamical studies later.

\subsection{Regularity and curvature invariants}
To show the regularity of this metric one may calculate several scalar curvature invariants. We focus here on the Ricci scalar $R$, the square of the traceless Ricci tensor $S^2 = (S{}_{\mu\nu})^2$, as well as the square of the Weyl tensor $C^2 = (C{}_{\mu\nu\rho\sigma})^2$ and the Kretschmann scalar $K = (R_{\mu\nu\rho\sigma})^2$. These quantities are related to each other via
\begin{align}
K = C^2 + 2 S^2 + \frac16 R^2 \, .
\end{align}
Their general expressions are quite cumbersome, so we will not show their explicit values here; at $r=0$ they take the following simpler form:
\begin{align}
R &= \frac{3\sqrt{\pi}M+ (6-2\pi)\ell}{3\pi\ell^3} - \frac{7\sqrt{\pi}M+2(10+\pi)\ell}{60\pi\ell^5} r^2 + \mathcal{O}(r^4) \, , \\
S^2 &= \frac{9\pi M^2+4(3-2\pi)\sqrt{\pi}M\ell + 2(18+\pi^2)\ell^2}{36\pi^2\ell^6} \nonumber \\
&\hspace{20pt} - \frac{63\pi M^2 + 2(51-13\pi)\sqrt{\pi}M\ell + 4[90 + (15+\pi)\pi]\ell^2}{1080\pi^2\ell^8} r^2 + \mathcal{O}(r^4) \, , \\
C^2 &= \frac{[3\sqrt{\pi}M-(6+\pi)\ell]^2}{27\pi^2\ell^6} - \frac{(27\sqrt{\pi}M-4(5+\pi)\ell)[3\sqrt{\pi}M-(6+\pi)\ell]}{270\pi^2\ell^8} r^2 + \mathcal{O}(r^4) \, , \\
K &= \frac{9\pi M^2 - 8 \pi^{3/2}M\ell+2(18+\pi^2)\ell^2}{9\pi^2\ell^6} \nonumber \\
&\hspace{20pt}- \frac{123\pi M^2 - 8\sqrt{\pi}(15+7\pi)M\ell + 4(90+15\pi+\pi^2)\ell^2 }{270\pi^2\ell^8} r^2 + \mathcal{O}(r^4) \, . 
\end{align}
Somewhat cumbersome expressions aside, it is clear that the scalar curvature at $r=0$ is positive for large masses $M > (2\pi-6)\ell/(3\sqrt{\pi})$, consistent with our previous estimate $2M > \sqrt{\pi}\ell$. Moreover, the invariants are all manifestly finite as well as regular at $r=0$, since no linear terms in $r$ appear.

\subsection{Limiting curvature condition}
However, the behavior of the invariants at $r=0$ is not bounded by a universal constant. Demanding that the curvature scales at most Planckian for typical astrophysical black holes,
\begin{align}
\mathcal{R} \sim \frac{GM_\odot}{c^2\ell^3} \lesssim \frac{1}{\ell_p^2} \, ,
\end{align}
gives the constraint that $\ell \gtrsim 10^{-22}\,\text{m}$, which is thirteen orders of magnitude larger than the Planck scale. Using this as a reference value, we can now estimate the order of magnitude of deviations from the Schwarzschild black hole outside the horizon of an astrophysical black hole, given by
\begin{align}
e^{-GM_\odot^2/(c^2\ell^2)} \approx e^{-10^{50}} \approx 0 \, .
\end{align}
This is to be compared to the case of polynomial non-singular black holes, where deviations are equal to simple powers of $c^2\ell/(GM_\odot) \sim 10^{-25}$.

\subsection{Effective energy-momentum tensor and energy conditions}

While the singularity-ridden black hole solutions of general relativity are vacuum solutions of the field equations, regular black hole models as the one presented in this paper do not solve these field equations. This is to be expected, since by virtue of the Birkhoff theorem the Schwarzschild solution is the unique spherically symmetric static vacuum solution of the field equations of general relativity. However, one may argue that in a UV-finite theory of gravity the field equations would deviate from those of general relativity, and hence it is not a substantial impediment that regular black hole models are no vacuum solutions.

One may certainly take the point of view that regular black hole models are supported by special types of matter, and then related the regularity properties (and deviations from the Schwarzschild metric) to the properties of this form of matter. For example, this has been achieved in the context of non-linear electrodynamics \cite{Ayon-Beato:1998hmi}, but this method does not work for all regular black holes, and hence this analysis is outside of the scope of the present paper. Alternatively, we may view $T{}_{\mu\nu} = G{}_{\mu\nu}/(8\pi)$ as the effective energy-momentum tensor of the proposed metric.

In that framework, we can now address energy conditions on the effective energy-momentum tensor. Since all energy conditions (dominant, weak, strong) imply the null energy condition, we opt to study the possible violation of the null energy condition as another indicator for the regularity of the proposed spacetime.

Following the discussion by Simpson and Visser \cite{Simpson:2018tsi}, we define the energy and density of the effective energy-momentum tensor as
\begin{align}
\rho = (-1) T{}^t{}_t \, , \quad
p_{||} = T{}^r{}_r \, , \quad
p_\perp = T{}^\theta{}_\theta = T{}^\varphi{}_\varphi \, .
\end{align}
The null energy condition is then equivalent to
\begin{align}
\rho + p_{||} \geq 0 \, , \quad
\rho + p_\perp \geq 0 \, .
\end{align}
Computing the effective energy-momentum tensor, the energy density $\rho$ is given by
\begin{align}
8\pi \rho &= \frac{\text{erf}\left( \frac{r}{2\ell} \right)^2 - 1}{r^2} + \frac{2M e^{-r^2/(4\ell^2)}}{\sqrt{\pi}\ell^3} - \frac{8M e^{-r^2/(4\ell^2)}}{\sqrt{\pi}r^2\ell} - \frac{5M e^{-r^2/(2\ell^2)}}{\pi\ell^2 \text{erf}\left( \frac{r}{2\ell} \right)^2} \nonumber \\
&\hspace{11pt}- \frac{r e^{-r^2/(4\ell^2)}}{\sqrt{\pi}\ell^3 \text{erf}\left( \frac{r}{2\ell} \right)} + \frac{8M e^{-r^2/(2\ell^2)}}{\pi r\ell^2 \text{erf}\left( \frac{r}{2\ell} \right)} + \frac{6 e^{-r^2/(4\ell^2)}}{\sqrt{\pi}r\ell \text{erf}\left( \frac{r}{2\ell} \right)} \, ,
\end{align}
whereas the parallel and transverse pressures $p_{||}$ and $p_\perp$ take the form
\begin{align}
8\pi p_{||} &= \frac{1 - \text{erf}\left( \frac{r}{2\ell} \right)^2}{r^2} + \frac{e^{-r^2/(2\ell^2)}}{\pi\ell^2 \text{erf}\left( \frac{r}{2\ell} \right)^2} - \frac{2 e^{-r^2/(4\ell^2)}}{\sqrt{\pi}r\ell \text{erf}\left( \frac{r}{2\ell} \right)} \, , \\
8\pi p_\perp &= \frac{e^{-r^2/(2\ell^2)} \left[ r - M \text{erf}\left( \frac{r}{2\ell} \right) \right]\left[ 4r\ell + \sqrt{\pi}(r^2-4\ell^2)\text{erf}\left( \frac{r}{2\ell} \right) e^{+r^2/(4\ell^2)} \right] }{2\pi r^2\ell^3 \text{erf}\left( \frac{r}{2\ell} \right)^2} \, .
\end{align}
Then one finds
\begin{align}
8\pi( \rho + p_{||} ) &= (-1)\frac{e^{-r^2/(2\ell^2)} \left[ r - 2M \text{erf}\left( \frac{r}{2\ell} \right) \right]\left[ 4r\ell + \sqrt{\pi}(r^2-4\ell^2)\text{erf}\left( \frac{r}{2\ell} \right) e^{+r^2/(4\ell^2)} \right] }{\pi r^2\ell^3 \text{erf}\left( \frac{r}{2\ell} \right)^2} \, , \\
8\pi ( \rho + p_\perp ) &= \frac{\text{erf}\left( \frac{r}{2\ell} \right)^2 - 1}{r^2} + \frac{3M e^{-r^2/(4\ell^2)}}{2\sqrt{\pi}\ell^3} - \frac{6M e^{-r^2/(4\ell^2)}}{\sqrt{\pi}r^2\ell} - \frac{3 e^{-r^2/(2\ell^2)}}{\pi\ell^2 \text{erf}\left( \frac{r}{2\ell} \right)^2} \nonumber \\
&\hspace{11pt}- \frac{r e^{-r^2/(4\ell^2)}}{\sqrt{\pi}\ell^3 \text{erf}\left( \frac{r}{2\ell} \right)} + \frac{6M e^{-r^2/(2\ell^2)}}{\pi r\ell^2 \text{erf}\left( \frac{r}{2\ell} \right)} + \frac{4 e^{-r^2/(4\ell^2)}}{\sqrt{\pi}r\ell \text{erf}\left( \frac{r}{2\ell} \right)} \, .
\end{align}
Recall that the black hole horizon is located at
\begin{align}
f(r_h) = 1 - \frac{2M}{r_h} \text{erf}\left( \frac{r_h}{2\ell} \right) = 0 \, ,
\end{align}
which implies that outside of the black hole, for $r > r_h$, one has
\begin{align}
r - 2M \text{erf}\left( \frac{r}{2\ell} \right) > 0 \, ,
\end{align}
implying that $\rho + p_{||} < 0$ outside of the black hole (assuming that $r > 2\ell$ which is always satisfied in the black hole exterior above the mass gap). Hence, the null energy condition is violated in the black hole exterior. The identical argument holds for \emph{inside} the black hole, as e.g.~Simpson and Visser point out \cite{Simpson:2018tsi}: Inside, $t$ and $r$ switch their places, and we define instead $\tilde{\rho} = (-1)T{}^r{}_r$ and $\tilde{p}_{||} = T{}^t{}_t$. Inside the horizon, the expression $r-M\text{erf}\left( \frac{r}{2\ell} \right)$ switches sign, but so does $\rho + p_{||} = (-1)T{}^t{}_t + T{}^r{}_r = -(\tilde{\rho} + \tilde{p}_{||})$. Hence, the null energy condition is identically violated past the outer horizon, in the black hole interior.

\begin{figure}[!htb]
\centering
\includegraphics[width=0.7\textwidth]{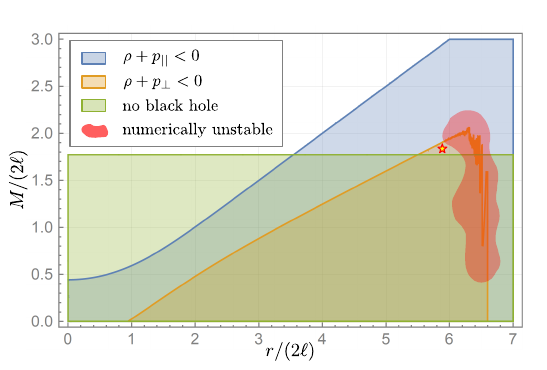}
\caption{Visual inspection of the violation of null energy conditions, expressed at fixed $\ell$ for various masses $M$ and radial distances $r$. The highlighted benchmark point is located at $\{r/(2\ell), M/(2\ell)\} = \{5.9, 1.8\}$ and violates the null energy condition as $8\pi G \ell^2 (\rho + p_{||}) = -1.921 \times 10^{-15} < 0$ and $8\pi G \ell^2 (\rho + p_\perp) = -1.379 \times 10^{-16} < 0$.}
\label{fig:4}
\end{figure}

Similarly, there exist values for which $\rho + p_\perp > 0$ is violated, but this expression is more cumbersome and hence difficult to study analytically. Hence, to avoid all ambiguities, we also numerically verified the above statements. To that end, it is useful to work in the dimensionless quantities $r/(2\ell)$ as well as $M/(2\ell)$; see Fig.~\ref{fig:4}.

In conclusion, this shows that the Gauss black hole, like all other known regular black hole models, violates one of the energy conditions in its vicinity as well as in its interior.

\subsection{Black hole thermodynamics}
To begin our considerations of the thermodynamic properties of the proposed metric, recall that the horizon of the black hole is defined implicitly via the transcendental equation $f(r_h) = 0$,
\begin{align}
\label{eq:horizon-radius}
r_h = 2M\text{erf}\left( \frac{r_h}{2\ell} \right) \, ,
\end{align}
provided that $M > M_0 = \sqrt{\pi}\ell/2$ (otherwise, no horizon exists). Since the geometry is static, one may apply standard Euclidean gravity techniques to extract the associated Hawking temperature as the periodicity of imaginary time \cite{Hartle:1976tp,Gibbons:1976ue,Gibbons:1976pt}, leading to
\begin{align}
T_H &= \frac{f'(r_h)}{4\pi} = \frac{2M}{4\pi r_h^2} \left[ \text{erf}\left( \frac{r_h}{2\ell} \right) - \frac{r_h}{\sqrt{\pi}\ell} \exp\left( -\frac{r_h^2}{4\ell^2} \right) \right]
\end{align}
However, the implicit nature of $r_h$ is obfuscating the physical significance of this expression. While it can be evaluated numerically (given the mass parameter $M$ as well as the regulator $\ell$), it is instructive to utilize Eq.~\eqref{eq:horizon-radius} to arrive at
\begin{align}
T_H = \frac{1}{4\pi r_h} - \frac{2M}{4\pi^{3/2} r_h \ell} \exp\left( -\frac{r_h^2}{4\ell^2} \right)
= \frac{1}{4\pi r_h} \left[ 1 - \frac{M}{M_0} \exp\left( -\frac{r_h^2}{4\ell^2} \right) \right] \, .
\end{align}
This form is interesting since it expresses the Hawking temperature to the would-be Schwarzschild temperature $1/(4\pi r_h)$, multiplied by a correction term involving both the regulator scale $\ell$ (which may be expected) as well as the mass gap $M_0$, which is somewhat less intuitive. Last, note that we can also recast Eq.~\eqref{eq:horizon-radius} into an expression for the mass. In that case, restoring the appearance of $\ell$, we arrive at a third expression for the temperature, 
\begin{align}
T_H = \frac{1}{4\pi r_h} \left[ 1 - \frac{1}{\sqrt{\pi}} \frac{r_h}{\ell} \frac{1}{\text{erf}\left( \frac{r_h}{2\ell} \right)} \exp\left( -\frac{r_h^2}{4\ell^2} \right) \right] \, .
\end{align}
Recalling the identity
\begin{align}
\lim_{\ell\rightarrow 0} \frac{e^{-r^2/(4\ell^2)}}{\sqrt{4\pi}\ell} = \delta{}^{(1)}(r)
\end{align}
one finds in the limit $\ell\rightarrow 0$ that
\begin{align}
T_H = \frac{1}{4\pi r_h} - \frac{M}{\pi} \delta{}^{(1)}(r_h) \, ,
\end{align}
where the last term vanishes identically since $r_h > 0$. This guarantees that in the absence of the regulator $\ell$ the black hole temperature coincides with the Schwarzschild case, as it must.

The black hole entropy, by similar Euclidean reasoning, is assumed to be given by the quarter of the area of the event horizon. While this step is particularly trivial in most regular black hole spacetimes (since the spherical part of the geometry is left unmodified) this is decidedly not the case in the present paper. Namely, one finds that the entropy is entirely regulator-independent and is directly given by the black hole mass. One computes
\begin{align}
S = \frac{A}{4} = \pi \frac{r_h^2}{\text{erf}\left(\frac{r_h}{2\ell}\right)^2} 
\overset{\eqref{eq:horizon-radius}}{=} 4 \pi M^2 \, .
\end{align}
This coincides with the Schwarzschild case, but it describes a black hole of a different composition. Importantly, the result follows again from the implicit relation \eqref{eq:horizon-radius}. This result is perhaps the most surprising one encountered in the context of the thermodynamic study of this black hole.

\begin{figure}[!htb]
\centering
\subfloat{ \includegraphics[width=0.5\textwidth]{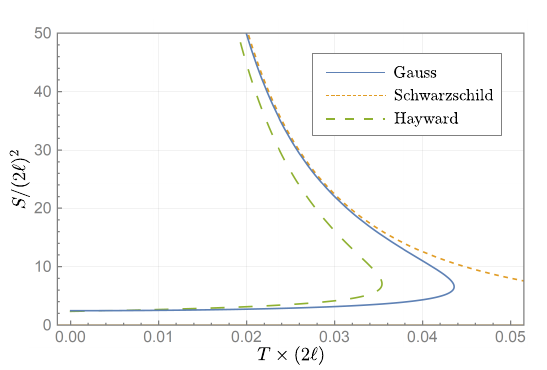} }
\subfloat{ \includegraphics[width=0.5\textwidth]{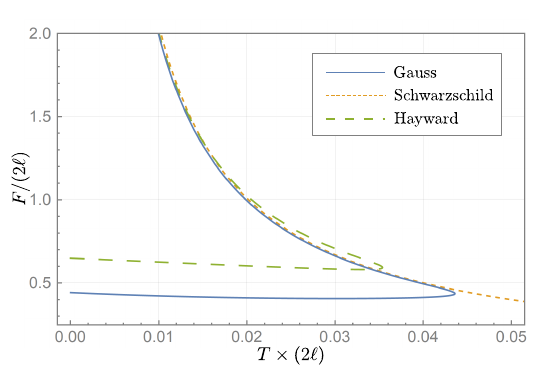} }
\caption{Dimensionless entropy (left) and dimensionless free energy (right) as a function of temperature. While a closed form of their functional relation is not available, the above diagrams have been generated parametrically in terms of the black hole horizon radius $r_h$. Qualitatively, the behavior of the Gauss and Hayward metric is similar, whereas they only approach the Schwarzschild behavior for large entropies or large free energies.}
\label{fig:5-6}
\end{figure}

Let us now address the thermodynamic stability of this metric by computing its specific heat. To begin with, we introduce a dimensionless temperature $\hat{T}_H = T_H \ell$ as well as a dimensionless horizon radius $\hat{r}_h = r_h/(2\ell)$, arriving at a compact expression for the temperature as a function,
\begin{align}
\hat{T}_H = \left( \frac{1}{4\pi \hat{r}_h} \right) \left[ 1 - \frac{2\hat{r}_h}{\sqrt{\pi}} \frac{e^{-\hat{r}_h^2}}{\text{erf}\left( \hat{r}_h \right)}  \right] \, ,
\end{align}
where we extracted the Schwarzschild prefactor in parentheses. Defining the specific heat in the usual manner, 
\begin{align}
C &= T \frac{\partial S}{\partial T} = T \frac{\partial S}{\partial r_h} \frac{\partial r_h}{\partial T} = \frac{T}{\frac{\partial T}{\partial r_h}} \frac{\partial S}{\partial r_h} \, ,
\end{align}
we can then express the dimensionless specific heat in terms of the horizon radius $r_h$ as follows:
\begin{align}
\hat{C} &= \frac{C}{(2\ell)^2} = \big( \! -2\pi \hat{r}_h^2 \big) \frac{ \sqrt{\pi} \, e^{\hat{r}_h^2} \, \text{erf}(\hat{r}_h) - 2 \hat{r}_h }{\text{erf}\left( \hat{r}_h \right)^2\left[ \pi \, e^{2\hat{r}_h^2} \, \text{erf}(\hat{r}_h)^2 - 4 \sqrt{\pi} \, \hat{r}_h^3 \, e^{\hat{r}_h^2} \, \text{erf}(\hat{r}_h) - 4 \hat{r}_h^2 \right]} \, .
\end{align}
Similar to above, we extracted the Schwarzschild value in leading parentheses. Last, we define the free energy by direct analogy to the Euclidean general relativity expression,
\begin{align}
F \equiv M - T \, S \, ,
\end{align}
and a related dimensionless free energy taking the form
\begin{align}
\hat{F} = \frac{F}{2\ell} = \left( \frac{\hat{r}_h}{4} \right) \frac{ 2\,\text{erf}(\hat{r}_h) - 1 + \frac{2 \hat{r}_h \, e^{-\hat{r}_h^2}}{\sqrt{\pi}\,\text{erf}(\hat{r}_h)} }{4 \, \text{erf}(\hat{r}_h)^2 } \, ,
\end{align}
where the prefactor is again the Schwarzschild value. Both the $ST$ and the $FT$ diagrams phase diagrams can now be constructed as parametrized plots of the dimensionless horizon radius $\hat{r}_h$. For definiteness, we included the expressions for Schwarzschild as well as those for the Hayward metric (where care has been taken that for the Gauss black hole and the Schwarzschild black hole we have $\hat{r} \geq 0$ and in the Hayward case we instead have $\hat{r}_h \geq \sqrt{3}/2$; see Fig.~\ref{fig:5-6}. Qualitatively, the behavior of the Gauss and Hayward case is comparable, namely, there exists a maximum temperature. The entropy is equal at the that maximum temperature for the Gauss and Hayward case, but the temperature itself is slightly larger in the Gauss case. Conversely, the free energy at maximum temperature is larger for the Hayward case.

\begin{figure}[!htb]
\centering
\includegraphics[width=0.7\textwidth]{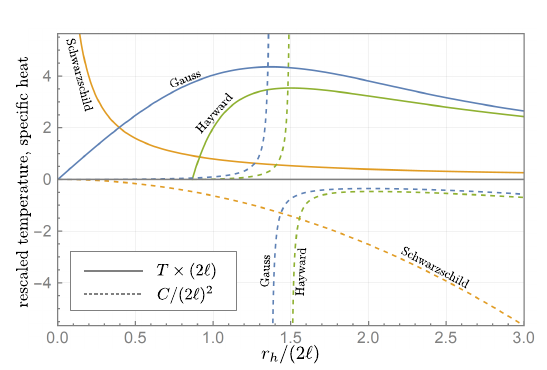}
\caption{We plot the dimensionless temperature and dimensionless specific heat (rescaled by convenient numericalt factors to fit them into one diagram) as a function of the dimensionless horizon radius. Both the Gauss and Hayward metric exhibit stable small black holes, whereas the specific heat diverges at an intermediate mass. Beyond that, like the Schwarzschild black hole, the resulting black hole configurations are unstable. The behavior of the Hawking temperature of both the Hayward and the Gauss black hole is similar, and approaches the Schwarzschild temperature case for black holes that are large compared to the regulator scale $\ell$.}
\label{fig:7}
\end{figure}

We would like to close this section by addressing the specific heat and the temperature of the Gauss black hole---for a graphical representation see Fig.~\ref{fig:7}. The specific heat is singular both in the Hayward and Gauss case in very similar fashion: for small black holes the specific heat is indicating stability; for large black holes, however, the specific heat turns out to be negative, similar to the Schwarzschild case, implying instability under Hawking radiation. However, an interesting consequence (that is not dissimilar to the fate of the Hayward metric) is that the decay under Hawking radiation will eventually terminate, once a sufficiently small mass is reached, resulting in a remnant. The discussion of this object, however, is outside of the scope of this paper, and may be addressed at a later stage.

\subsection{Properties of $r=0$, wormholes, and geodesic (in)completeness}
The location $r=0$ corresponds to $\tilde{r} = \sqrt{\pi}\ell$ and hence the metric is
\begin{align}
\left.\dd s^2\right|_{r=0} = \left(\frac{2M}{\sqrt{\pi}\ell} - 1\right)\dd t^2 + \pi\ell^2\dd\Omega^2 \, ,
\end{align}
which is nothing but a sphere of surface area $4\pi^2\ell^2$ factored with another spatial direction $t$, provided the mass parameter $M$ is large enough. It would be interesting to study the response of this ``throat'' to infalling matter. A radial null geodesic in a static metric with $-g_{tt} = g^{rr} = f(r)$ has a conserved quantity $E = f(r)\dot{t}$, where the dot denotes differentiation with respect to the affine parameter $\lambda$. Then,
\begin{align}
\dot{r}^2 = E^2 \, ,
\end{align}
implying that any radial geodesic can reach the surface of that sphere ($r=0$) at finite affine parameter, which in turn implies geodesic incompleteness \cite{Frolov:2011}; for an application to regular black holes see Ref.~\cite{Bambi:2016wdn}. However, this might not be a serious drawback since many regular black hole models are geodesically incomplete \cite{Carballo-Rubio:2019fnb}. In this particular case it seems that continuing the variable $r$ to the entire range of $\mathbb{R}$ would solve that issue and potentially give rise to a wormhole-type geometry; see also Simpson and Visser \cite{Simpson:2018tsi}. For that reason, even though from the outside the proposed Gauss metric appears as a black hole (and has been proposed as a candidate for a regular black hole metric), its properties under analytic continuation may deserve further scrutiny.

\section{Conclusions and outlook}

We have proposed a non-singular ``Gauss'' black hole from the principle of a non-local regulator that ``cuts out'' a piece of spacetime with radii less than the non-local regularization scale, providing a mechanism for the procedure described e.g.\ by Klinkhamer \cite{Klinkhamer:2013vva}. The presented geometry has several interesting features: first, it has no inner horizon. Second, its deviation from the Schwarzschild vacuum decreases exponentially fast, which---similar to the Dymnikova black hole \cite{Dymnikova:1992ux}---makes it a rather good approximative vacuum solution, and its effective energy-momentum tensor demonstrably violates the null energy condition. And third, perhaps most interestingly, it provides a mechanism to arrive at a regular spacetime. However, this model does not satisfy the limiting curvature condition, thereby placing a constraint of $\ell \gtrsim 10^{-22}\,\text{m}$ on the scale of non-locality, when applied to astrophysical black holes. Its thermodynamics resembles that of other regular black holes, with the interesting difference that at the critical mass $M_0$, below which the horizon ceases to exist, the horizon radius also approaches zero. The fact that the Gauss black hole becomes arbitrarily small at finite mass may have interesting applications in quantum gravity phenomenology.

\section*{Acknowledgements}

The author thanks Pablo A.~Cano (Leuven), Piero Nicolini (Frankfurt), and Dejan Stojkovic (Buffalo) for their comments and for pointing out additional references. He is moreover grateful to three anonymous reviewers, whose comments have improved this manuscript appreciably. The author would also like to thank Valeri Frolov (Edmonton) for helpful comments, and was supported by the National Science Foundation under grants PHY-1819575 and PHY-2112460 while employed at William \& Mary, VA, USA. During the substantial editing phase, the author acknowledges support as a Fellow of the Young Investigator Group Preparation Program at KIT, Germany. This program receives joint funding via the University of Excellence strategic fund at the Karlsruhe Institute of Technology (administered by the federal government of Germany) and the Ministry of Science, Research and Arts of Baden-W\"urttemberg (Germany).

\begin{singlespace}

\end{singlespace}

\end{document}